\documentclass[12pt]{article}
\usepackage{epsf}
\title{ {\bf The $\tau\rightarrow \mu \, \bar{\nu_i} \, \nu_i$ decay
in the Randall Sundrum background with localized $U(1)_Y$ gauge
boson}}
\author{\vspace{1cm}\\
        {\bf E. O. Iltan,}
        \thanks{E-mail address:
        eiltan@newton.physics.metu.edu.tr}
{\bf H. Sundu}
        \thanks{E-mail address:
        sundu@metu.edu.tr}
 \\
        Physics Department, Middle East Technical University \\
        Ankara, Turkey\\}
\date{}

\begin{document}
\setlength{\baselineskip}{24pt}
\maketitle
\setlength{\baselineskip}{7mm}
\begin{abstract}
We study the effects of localization of the $U(1)_Y$ gauge boson
around the visible brane and the contributions of the KK modes of
Z bosons on the BR of the LFV $\tau\rightarrow \mu \, \bar{\nu_i}
\, \nu_i$ decay. We observe that the BR is sensitive to the amount
of localization of Z boson in the bulk of the Randall Sundrum
background.
\end{abstract}
\thispagestyle{empty}
\newpage
\setcounter{page}{1}
\section{Introduction}
Lepton flavor violating (LFV) interactions are rich from the
theoretical point of view since they exist at the loop level and
the measurable quantities of these decays carry considerable
information about the free parameters of the model used. In
addition to this, they are clean theoretically because they are
free from the nonperturbative effects. The experimental work done
stimulate the theoretical studies on LFV decays. The
$\mu\rightarrow e\gamma$ and $\tau\rightarrow \mu \gamma$
processes are among the LFV decays and the experimental the
current limits for their branching ratios (BR) have been predicted
as $1.2\times 10^{-11}$ \cite{Brooks} and $1.1\times 10^{-6}$
\cite{Ahmed}, respectively.

In the framework of the standard model (SM) the theoretical values
of the BRs of LFV interactions are extremely small. Therefore, one
search new models beyond the SM to enhance these numerical values
and the two Higgs doublet model (2HDM), with flavor changing
neural currents (FCNC) at tree level, is one of the candidate. In
this model, the LFV interactions are induced by the internal
neutral Higgs bosons $h^0$ and $A^0$ with the help of the Yukawa
couplings, appearing as free parameters, which can be determined
by the experimental data.

The present work is devoted to the analysis of $\tau\rightarrow
\mu \,\bar{\nu_i}\, \nu_i$, $i=e,\mu,\tau$ decay in the 2HDM with
the inclusion of a single extra dimension, respecting the Randall
Sundrum model (see \cite{iltHay} and \cite{iltExtr} for the
calculation of the BR of the same decay in the 2HDM without and
with flat extra dimensions). Here, the LFV transition
$\tau\rightarrow \mu$ is driven by the internal scalar bosons
$h^0$ and $A^0$ and the internal Z boson connects this transition
to the $\bar{\nu}\nu$ output (see Fig.
\ref{fig1ver})\footnote{Here, we respect the assumption that the
Cabibbo-Kobayashi-Maskawa (CKM) type matrix in the leptonic sector
doest not exist and the charged flavor changing (FC) interactions
vanish.}.

The hierarchy problem between weak and Planck scales could be
explained by introducing the extra dimensions. The model
introduced by Randall and Sundrum \cite{Rs1, Rs2} (the RS1 model)
is related to the non-factorizable geometry where the gravity is
localized in a 4D brane, so called Planck (hidden) brane, which is
away from another 4D brane, which is the visible (TeV) brane where
we live. In this scenario, the hierarchy is generated by the
warped factor, which is an exponential function of the
compactified radius in the extra dimension. There are other
scenarios based on the RS1 background in the literature
\cite{Goldberger}-\cite{Batell}. In \cite{Pamoral,Hewett} the
behavior of $U(1)$ gauge boson, accessible to the extra dimension
in the RS1 background, has been studied and it was observed that
the massless mode was not localized in the extra dimension.
Furthermore, it was obtained that the KK excitations have large
couplings to boundaries and this was not a phenomenologically
favorable scenario since, for a perturbative theory, it would have
been necessary to push the visible scale to energies greater than
TeV. To have a zero mode localized in the bulk and to get small
couplings of KK modes with the boundaries, the idea of brane
localized mass terms has been considered for scalar fields
\cite{Pamoral2}. These terms could change the boundary conditions
to get a zero mode localized solution. In \cite{Pamoral2,
Grossman}, the bulk fermions were considered in the RS1
background. \cite{Kogan} was devoted to a extensive work on the
bulk fields in various multi-brane models. In \cite{Batell}, the
$U(1)$ gauge field with bulk and boundary mass terms were taken
into account and only the $U(1)_Y$ gauge field was considered as
localized in the bulk.

In this work, we study the effects of localization of the $U(1)_Y$
gauge boson around the visible brane and analyze the contributions
of the KK modes of Z bosons on the BRs of the LFV $\tau\rightarrow
\mu \, \bar{\nu_i} \, \nu_i$ decay, by following the idea that the
$U(1)_Y$ gauge field is accessible to the extra dimension in the
RS1 background and the other particles, including the new Higgs
doublet lie on the visible brane.

The paper is organized as follows: In Section 2, we present the
theoretical expression for the decay width of the LFV decay
$\tau\rightarrow \mu \,\bar{\nu_i}\, \nu_i$, $i=e,\mu,\tau$, in
the framework of the 2HDM by considering the non zero localization
of the $U(1)_Y$ gauge in the RS1 background. Section 3 is devoted
to discussion and our conclusions. In Appendix A, we present the
model construction. Appendix B is devoted to the explicit
expressions of the functions appearing in the general effective
vertex for the interaction of off-shell Z-boson with a fermionic
current.
\section{The effect of the localized $U(1)_Y$ gauge
boson on the $\tau\rightarrow \mu \, \bar{\nu_i} \, \nu_i$ decay
in the Randall Sundrum background}
LFV  $\tau\rightarrow \mu \, \bar{\nu_i} \, \nu_i$ decay exists at
least in the one loop level and, therefore, the physical
quantities, like the BR, contain rich information about the model
used and the free parameters existing. In the version of the 2HDM
with the FCNCs at tree level, the LFV interactions exist with
larger BRs, compared to ones obtained in the extended SM with
massive neutrinos. These decays are driven by the Yukawa
interaction lagrangian and the strength of the interaction is
controlled by the new Yukawa couplings. The additional effects
coming from the possible extra dimension(s) bring new
contributions to the BRs of these decays and, in the present work,
we study these effects by assuming that the $U(1)_Y$ gauge boson
is accessible to the extra dimension and is localized on the
visible brane, in the RS1 background.

The RS1 model is a higher dimensional scenario that is based on a
non-factorizable geometry and the hierarchy is generated by the
warped factor, which is an exponential function of the
compactified radius in the extra dimension. This model is based on
the idea that the gravity is localized on the so called hidden
brane, which is one of the boundary of the $S^1/Z_2$ orbifold
extra dimension, and  extended into the bulk with varying
strength, however, the SM fields  live in another brane, the so
called visible brane, which is the other boundary of the extra
dimension. In this scenario, the 5D cosmological constant is non
vanishing and both branes have equal and opposite tensions so that
the low energy effective theory has flat 4D spacetime. The metric
of such 5D world reads
\begin{eqnarray}
ds^2=e^{-2\,k\,y}\,\eta_{\mu\nu}\,dx^\mu\,dx^\nu+dy^2\, ,
\label{metric1}
\end{eqnarray}
where $k$ is the bulk curvature constant, $R$ is the
compactification radius, $y=R\,|\theta|,\, 0\leq |\theta|\leq \pi$
and $e^{-k\,R\,|\theta|}$ is the warp factor which causes that all
the mass terms are rescaled on the visible brane for $\theta=\pi$.
With the rough estimate, $kR\sim 11-12$, all mass terms are bring
down to the TeV scale.

\begin{figure}[htb]
\vskip -0.5truein \centering \epsfxsize=6.0in
\leavevmode\epsffile{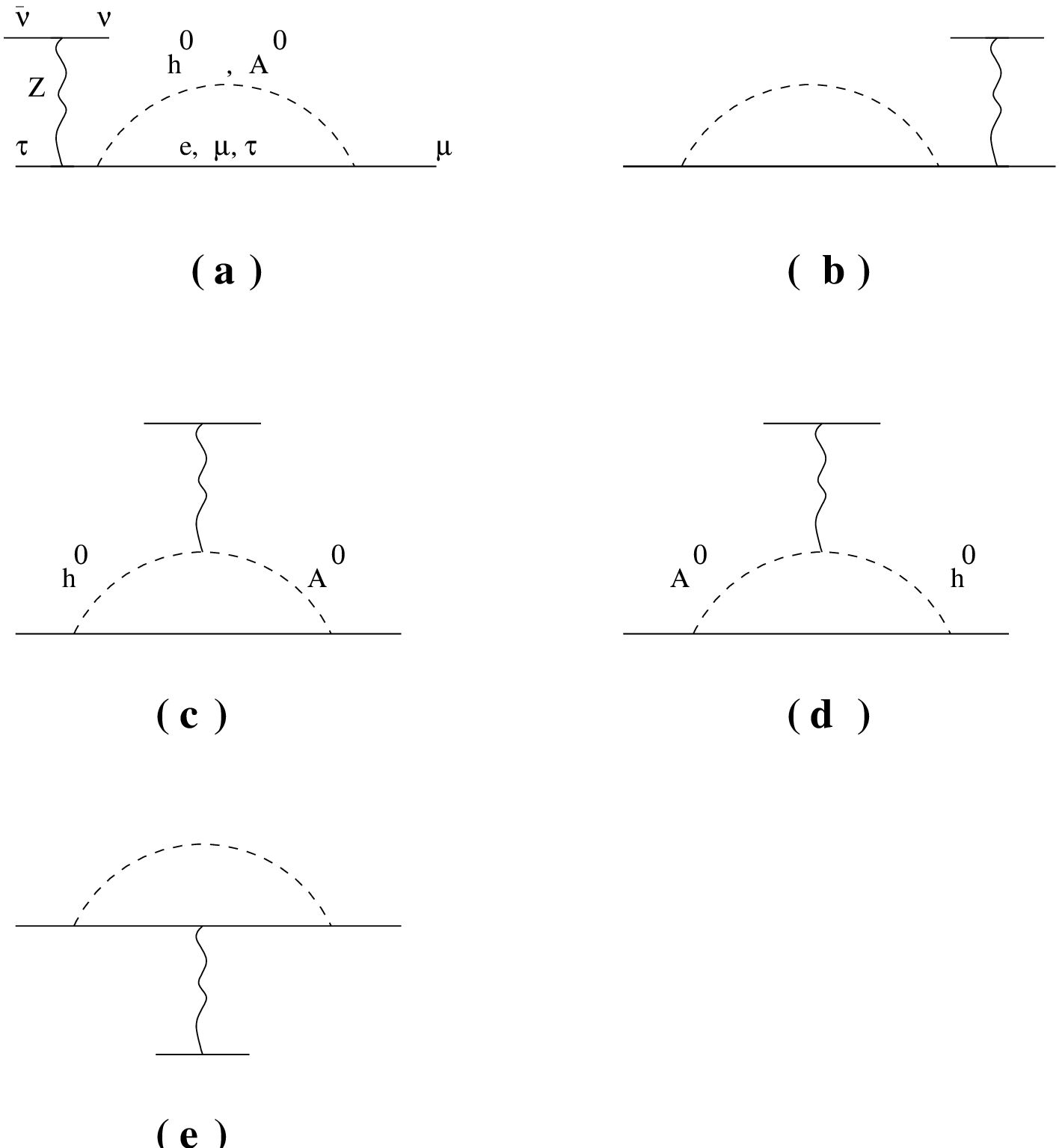} \vskip 0.5truein \caption[]{One loop
diagrams contribute to $\tau\rightarrow \mu \,\bar{\nu_i}\,
\nu_i$, $i=e,\mu,\tau$ decay due to the neutral Higgs bosons $h_0$
and $A_0$ in the 2HDM. Solid lines represent leptons and
neutrinos, curly (dashed) lines represent the virtual $Z$ boson
and its KK modes ($h_0$ and $A_0$ fields).} \label{fig1ver}
\end{figure}

The LFV  $\tau\rightarrow \mu \, \bar{\nu_i} \, \nu_i$ decay is
induced by $\tau\rightarrow \mu  Z^*$ transition and  $Z^* \,
\rightarrow \bar{\nu_i} \, \nu_i$ process (see Fig.
\ref{fig1ver}). The $\tau\rightarrow \mu Z^*$ transition, which
needs the FCNC at tree level, is driven by the internal new
neutral Higgs bosons $h^0$ and $A^0$, which are living on the
visible  brane. Now, we present the Yukawa interaction, which is
responsible for the $\tau\rightarrow \mu$ transition of the decay,

\begin{eqnarray}
{\cal{S}}_{Y}= \int d^5x \sqrt{-g} \,\Bigg(
\xi^{E}\,\bar{l}_{i L} \phi_{2} E_{j R} + h.c. \Bigg)\,
\delta(y-\pi R) \,\,\, , \label{lagrangian1}
\end{eqnarray}
where $L$ and $R$ denote chiral projections $L(R)=1/2(1\mp
\gamma_5)$, $\phi_{2}$ is the new scalar doublet, $l_{i L}$ ($E_{j
R}$) are lepton doublets (singlets), $\xi^{E}_{ij}$ \footnote{In
the following, we replace $\xi^{E}$ with $\xi^{E}_{N}$ where "N"
denotes the word "neutral".}, with family indices $i,j$ , are the
Yukawa couplings which induce the FV interactions in the lepton
sector and $g$ is the determinant of the metric\footnote{ Notice
that the term $\sqrt{-g}=e^{-4\,k\,y}$ is embedded into the
redefinitions of the fields on the visible brane for $y=\pi\,R$,
namely, they are warped as $\phi_{2}\rightarrow
e^{k\,\pi\,R}\,\phi^{warp}_{2}$, $l_i\rightarrow
e^{3\,k\,\pi\,R/2}\,l^{warp}_i$ and in the following we use warped
fields without the $warp$ index.} (see eq. (\ref{metric1})). We
assume that the Higgs doublet $\phi_1$ has non-zero vacuum
expectation value to ensure the ordinary masses of the gauge
fields and the fermions, however, the second doublet has no vacuum
expectation value, namely, we choose the doublets $\phi_{1}$ and
$\phi_{2}$ and their vacuum expectation values as
\begin{eqnarray}
\phi_{1}=\frac{1}{\sqrt{2}}\left[\left(\begin{array}{c c}
0\\v+H^{0}\end{array}\right)\; + \left(\begin{array}{c c} \sqrt{2}
\chi^{+}\\ i \chi^{0}\end{array}\right) \right]\, ;
\phi_{2}=\frac{1}{\sqrt{2}}\left(\begin{array}{c c} \sqrt{2}
H^{+}\\ H_1+i H_2 \end{array}\right) \,\, , \label{choice}
\end{eqnarray}
and
\begin{eqnarray}
<\phi_{1}>=\frac{1}{\sqrt{2}}\left(\begin{array}{c c}
0\\v\end{array}\right) \,  \, ; <\phi_{2}>=0 \,\, .\label{choice2}
\end{eqnarray}
This choice ensures that the mixing between neutral scalar Higgs
bosons is switched off and it would be possible to separate the
particle spectrum so that the SM particles are collected in the
first doublet and the new particles in the second
one\footnote{Here $H^1$ ($H^2$) is the well known mass eigenstate
$h^0$ ($A^0$).}.

In our analysis, we further assume that the $U(1)_Y$ gauge boson
is accessible to the extra dimension and the other
particles,including $SU(2)_L$ gauge bosons are confined on the
visible brane. This leads to an effective localization of Z boson
in the bulk of RS1 background. Therefore, the physical quantities
related to the decay studied get additional contributions coming
from  the internal Z boson and its KK modes (see \cite{Batell} for
details and Appendix A for the summary of the model construction).

Now, we present the general effective vertex for the interaction
of off-shell Z-boson with a fermionic current
\begin{equation}
\Gamma^{(REN)}_\mu (\tau\rightarrow \mu Z^*)=  f_1 \, \gamma_\mu +
f_2  \,\gamma_\mu \gamma_5+f_3\, \sigma_{\mu\nu}
k^\nu+f_4\,\sigma_{\mu\nu}\gamma_5 k^\nu  \, ,\label{GammaRen2}
\end{equation}
where $k$ is the momentum transfer, $k^2=(p-p')^2$, $p$
($p^{\prime}$) is the four momentum vector of incoming (outgoing)
lepton and the explicit expressions for the functions $f_1$,
$f_2$, $f_3$ and $f_4$ are given in Appendix B. The matrix element
$M$ of the $\tau\rightarrow \mu \, \bar{\nu_i}\, \nu_i$ process is
obtained by connecting  the $\tau\rightarrow \mu $ transition to
the $\bar{\nu_i} \nu_i$ pair with the internal Z boson and its KK
modes (see Fig. \ref{fig1ver}). For the $n^{th}$ internal Z KK
mode contribution, the coupling $g$ in $f_i, i=1,...4$, and at the
$Z^{(n)} \bar{\nu_i} \nu_i$ vertex is replaced by $g^{(n)}\sim
\frac{g}{\sqrt{\alpha}}$ and the $Z$ boson KK $n$ mode propagator
is obtained by using the KK mode mass $m_n$ (see eq. (\ref{mn})).
Using the matrix element $M$, the decay width $\Gamma$ of the
decay under consideration can be obtained in the $\tau$ lepton
rest frame with the help of the well known expression
\begin{equation}
d\Gamma=\frac{(2\, \pi)^4}{2\, m_\tau} \, |M|^2\,\delta^4
(p-\sum_{i=1}^3 p_i)\,\prod_{i=1}^3\,\frac{d^3p_i}{(2 \pi)^3 2
E_i} \,
 ,
\label{DecWidth}
\end{equation}
where $p$ ($p_i$, i=1,2,3) is four momentum vector of $\tau$
lepton ($\mu$ lepton, incoming $\nu$, outgoing $\nu$).
\section{Discussion}
The BRs of the LFV decays are negligible in the SM, including
non-zero neutrino masses. The extension of the Higgs sector brings
new LFV vertices with the help of the new Higgs scalars. 2HDM with
FCNC at tree level is the most primitive model in order to turn on
the LFV interactions and the $\tau\rightarrow \mu \, \bar{\nu_i}
\, \nu_i$ is induced by the LFV $\tau\rightarrow \mu$ transition
which exists at least at one loop level in this model. The Yukawa
couplings $\bar{\xi}^E_{N,ij}, i,j=e, \mu, \tau$ are the essential
parameters driving the lepton flavor violation and they are free
parameters which should be fixed by present and forthcoming
experiments.\footnote{The dimensionfull Yukawa couplings
$\bar{\xi}^{E}_{N,ij}$  are defined as
$\xi^{E}_{N,ij}=\sqrt{\frac{4\,G_F}{\sqrt{2}}}\,
\bar{\xi}^{E}_{N,ij}$.}
In our calculations, we assume that the Yukawa couplings
$\bar{\xi}^{E}_{N,ij}$ are symmetric with respect to the indices
$i$ and $j$ and take $\bar{\xi}^{E}_{N,ij},\, i,j=e,\mu $, as
small compared to $\bar{\xi}^{E}_{N,\tau\, i}\, i=e,\mu,\tau$
since we consider that the strengths of these couplings are
related with the masses of leptons denoted by the indices of them.
Therefore, we take only the $\tau$ lepton as an internal one and
we choose the couplings $\bar{\xi}_{N,\tau\tau}^E$ and
$\bar{\xi}_{N,\tau\mu}^E$ as non-zero. The upper limit of the
coupling $\bar{\xi}^{E}_{N,\tau \mu}$ has been estimated as $30\,
GeV$ (see \cite{Iltananomuon} and references therein) by assuming
that the new physics effects can not exceed experimental
uncertainty $10^{-9}$ in the measurement of the muon anomalous
magnetic moment. In our numerical calculation we choose
$\bar{\xi}^{E}_{N,\tau \mu}=1\,GeV$ by respecting this upper
limit. Since there is no restriction for the Yukawa coupling
$\bar{\xi}^{E}_{N,\tau \tau}$, the numerical values we use are
greater than $\bar{\xi}^{E}_{N,\tau \mu}$.

The addition of a single extra spatial dimension brings new
contributions to the BR of the decay under consideration and the
source of these contributions are the KK excitations of the
particles, which live in the bulk. Here,  we study the BR of the
LFV process  $\tau\rightarrow \mu \, \bar{\nu_i} \, \nu_i$ in the
framework of the 2HDM, including the extra dimension effects in
the RS1 scenario. The RS1 model is an alternative scenario to
solve the well known hierarchy problem. It is based on the
assumption that the extra dimension is compactified into $S^1/Z_2$
orbifold with two 4D surfaces (branes) at the boundaries in 5D
world and the extra dimensional bulk is populated by gravity,
which is localized on the hidden brane. Furthermore, the SM
particles live on the so called visible brane. However, in our
case, we follow the idea \cite{Batell} that the $U(1)_Y$ gauge
field is accessible to the extra dimension in the RS1 background
and the other particles, including the new Higgs doublet, lie on
the visible brane\footnote{Here we assume that the corresponding
gauge field make small contribution to the bulk energy density not
to disturb the solution of the Einstein's equations.}. With the
help of the boundary mass term (see eq. (\ref{Action3})) it is
possible to get zero the mode term and this mode is localized
around the visible brane with the special choice of the
parameters, $a$ and $\alpha$, related to the bulk and boundary
mass terms (eq. (\ref{alfaa})). If the parameter $\alpha$
vanishes, namely, there is no boundary mass term, KK mode coupling
to fermions becomes almost one order larger compared to the zero
mode one and, in order to obtain a perturbative theory, it would
be necessary to push the visible scale to energies greater than
TeV (see the discussion given in \cite{Pamoral,Hewett}). This is
not a phenomenologically favorable scenario. For $\alpha> 0$ the
zero mode is localized around the visible brane and the KK mode
fermion coupling is small enough for $\alpha> 1$ to obey the
phenomenology. Notice that one gets the original RS1 model for
infinitely large $\alpha$.

In the present work, we study the effects of localization of the
$U(1)_Y$ gauge boson and the contributions of the KK modes of Z
bosons on the BRs of the LFV $\tau\rightarrow \mu \, \bar{\nu_i}
\, \nu_i$ decay. Throughout our calculations we use the input
values given in Table (\ref{input}).
\begin{table}[h]
        \begin{center}
        \begin{tabular}{|l|l|}
        \hline
        \multicolumn{1}{|c|}{Parameter} &
                \multicolumn{1}{|c|}{Value}     \\
        \hline \hline
        $m_{\mu}$                   & $0.106$ (GeV) \\
        $m_{\tau}$                  & $1.78$ (GeV) \\
        $m_{h^0}$           & $100$   (GeV)  \\
        $m_{A^0}$           & $200$   (GeV)  \\
        $G_F$             & $1.16637 10^{-5} (GeV^{-2})$  \\
        \hline
        \end{tabular}
        \end{center}
\caption{The values of the input parameters used in the numerical
          calculations.}
\label{input}
\end{table}

\begin{figure}[htb]
\vskip -3.0truein \centering \epsfxsize=6.8in
\leavevmode\epsffile{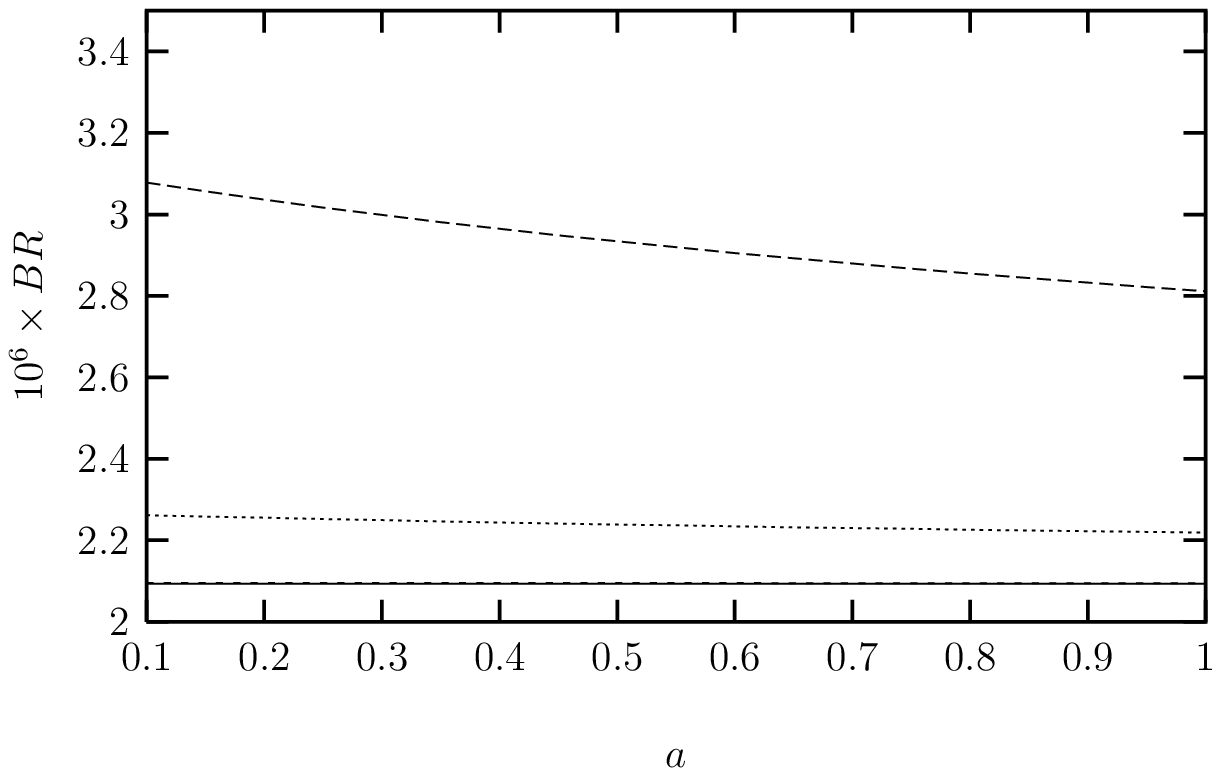} \vskip -3.0truein \caption[]{The
parameter $a$ dependence of the BR of the LFV $\tau\rightarrow \mu
\, \bar{\nu_i} \, \nu_i$ decay for
$\bar{\xi}^{E}_{N,\tau\tau}=10\, GeV$ and $\bar{\xi}^{E}_{N,\tau
\mu}=1\, GeV$. The solid (dashed-small dashed-dotted) line
represents the BR without extra dimension (with extra dimension
for $k R=12,\, k=10^{18}\,GeV-k R=11,\, k=10^{18}\,GeV-k R=11,\,
k=10^{17}\,GeV$).} \label{BRa}
\end{figure}

In Fig. \ref{BRa}, we present the parameter $a$ dependence of the
BR of the  LFV $\tau\rightarrow \mu \, \bar{\nu_i} \, \nu_i$ decay
for $\bar{\xi}^{E}_{N,\tau\tau}=10\, GeV$ and
$\bar{\xi}^{E}_{N,\tau \mu}=1\, GeV$. The solid (dashed-small
dashed-dotted) line represents the BR without extra dimension
(with extra dimension for $k R=12,\, k=10^{18}\,GeV-k R=11,\,
k=10^{18}\,GeV-k R=11,\, k=10^{17}\,GeV$). This figure shows that
the BR is at the order of the magnitude of $10^{-6}$ for the free
parameters used and it decreases with the increasing values of
$a$. This is due to the fact that the interaction coupling of Z
boson KK modes to the fermions becomes weak with the increasing
values of the zero mode localization parameter $\alpha$ and the
increase in KK mode masses results in a suppression in the KK mode
contribution. For the limit $a\rightarrow \infty$, this coupling
vanishes, the zero mode is confined in the visible brane and we
reach the numerical values of BR for the original RS1 model, where
all 2HDM particles are confined in the visible brane. On the other
hand, we observe that for decreasing values of $k R$ the BR
becomes smaller because of the weaker coupling $g^{(n)}$.

\begin{figure}[htb]
\vskip -3.0truein \centering \epsfxsize=6.8in
\leavevmode\epsffile{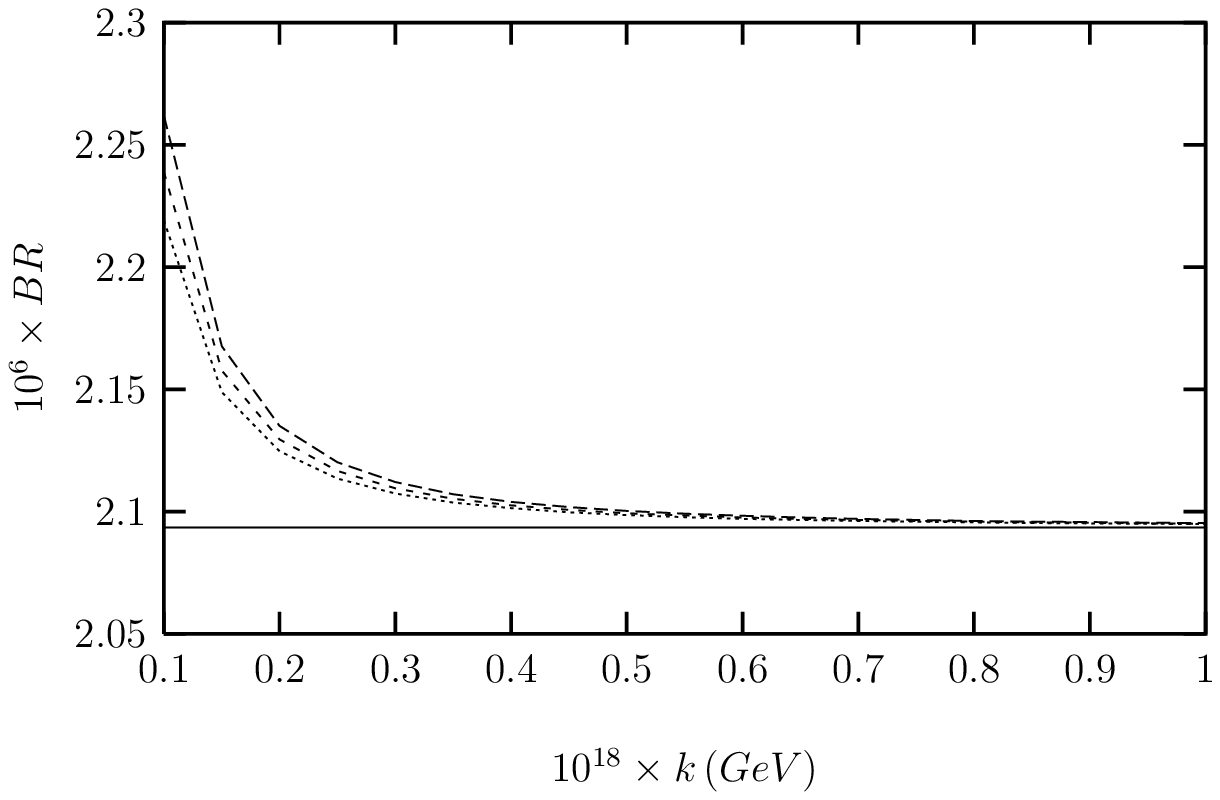} \vskip -3.0truein \caption[]{The
parameter $k$ dependence of the BR of the LFV $\tau\rightarrow \mu
\, \bar{\nu_i} \, \nu_i$ decay for
$\bar{\xi}^{E}_{N,\tau\tau}=10\, GeV$ and $\bar{\xi}^{E}_{N,\tau
\mu}=1\, GeV$ and $kR=11$. The solid (dashed-small dashed-dotted)
line represents the BR without extra dimension (with extra
dimension for $a=0.1-0.5-1.0$).} \label{BRk}
\end{figure}

Fig. \ref{BRk} is devoted to the parameter $k$ dependence of the
BR($\tau\rightarrow \mu \, \bar{\nu_i} \, \nu_i$)  for
$\bar{\xi}^{E}_{N,\tau\tau}=10\, GeV$ and $\bar{\xi}^{E}_{N,\tau
\mu}=1\, GeV$ and $kR=11$. The solid (dashed-small dashed-dotted)
line represents the BR without extra dimension (with extra
dimension for $a=0.1-0.5-1.0$). It is observed that the BR
decreases with the increasing values of $k$. Here the KK mode
masses increases with $k$ for fixed $kR$ and, as a result, the BR
is suppressed and reaches to the one which is obtained without
extra dimension.

\begin{figure}[htb]
\vskip -3.0truein \centering \epsfxsize=6.8in
\leavevmode\epsffile{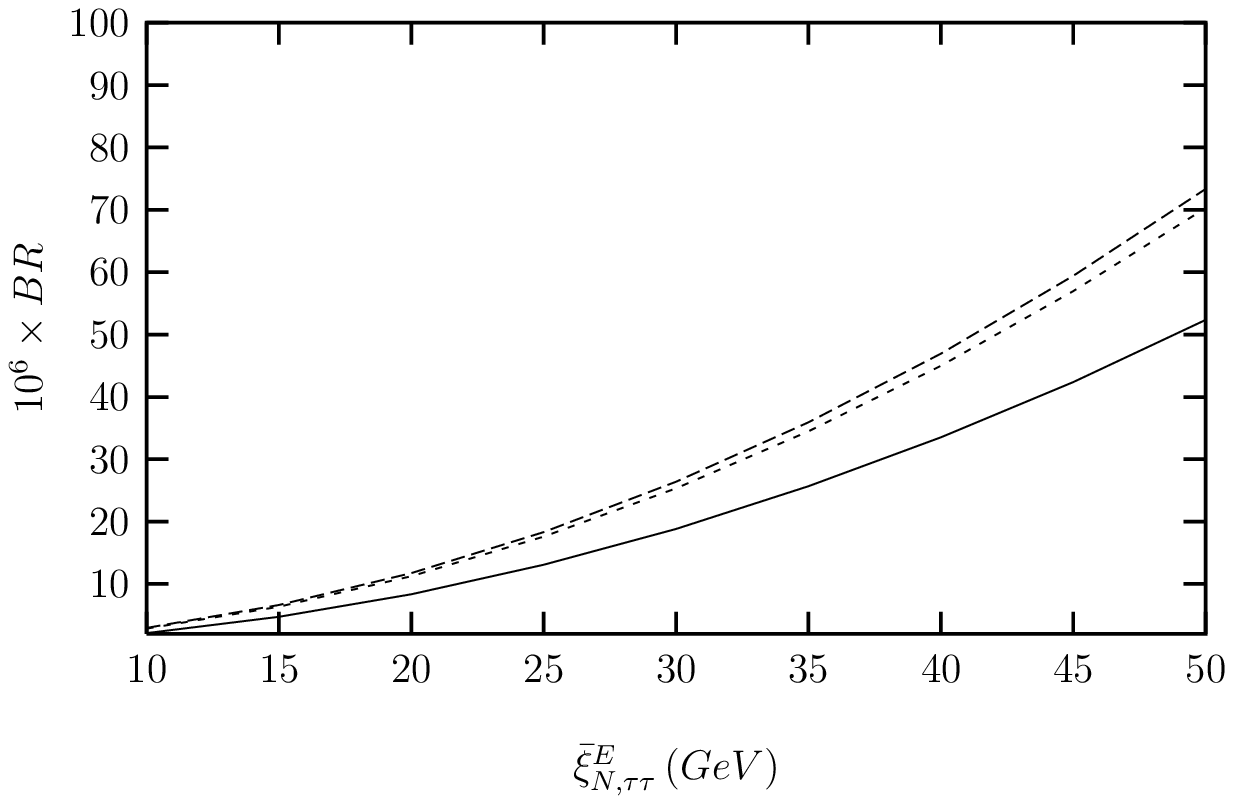} \vskip -3.0truein \caption[]{
$\bar{\xi}^{E}_{N,\tau\tau}$ dependence of the BR of the LFV
$\tau\rightarrow \mu \, \bar{\nu_i} \, \nu_i$ decay for
$\bar{\xi}^{E}_{N,\tau \mu}=1\, GeV$, $kR=12$ and $k=10^{18}\,
GeV$. The solid (dashed-small dashed) line represents the BR
without extra dimension (with extra dimension for $a=0.5-1.0$).}
\label{BRksi}
\end{figure}

Finally, for completeness, we study the Yukawa coupling dependence
of the BR for different values of the parameter $a$. Fig.
\ref{BRksi} represents $\bar{\xi}^{E}_{N,\tau\tau}$ dependence of
the BR of the LFV $\tau\rightarrow \mu \, \bar{\nu_i} \, \nu_i$
decay, for $\bar{\xi}^{E}_{N,\tau \mu}=1\, GeV$, $kR=12$ and
$k=10^{18}\, GeV$. The solid (dashed-small dashed) line represents
the BR without extra dimension (with extra dimension for
$a=0.5-1.0$). The BR is sensitive to the Yukawa couplings as it
should be and the increase in the parameter $a$ pushes the
numerical value of the BR to the one obtained without extra
dimension because of strong localization of the Z field around the
visible brane. This is case that the original RS1 model is
reached, where the extra dimension effects are switched off for
the process under consideration.

As a result, the BR of the LFV $\tau\rightarrow \mu \, \bar{\nu_i}
\, \nu_i$ decay is sensitive to the parameter $a$ and it decreases
with the increasing values of $a$. In addition to this, the BR
decreases with the increasing values of $k$ for fixed $kR$, since
the masses of KK modes are proportional to $k$. The sensitivity of
the BR($\tau\rightarrow \mu \, \bar{\nu_i} \, \nu_i$) to the extra
dimension effects in RS1 model is informative and the more
accurate future experimental measurement of this decay can ensure
a possible test for the existence of the model used and the
determination of the signals coming from the extra dimensions.
\newpage
\appendix

\vskip0.8cm \noindent \centerline{\Large \bf Appendix} \vskip0.4cm
\noindent

\section{Model Construction}
Now, we would like to summarize the model construction, following
the work \cite{Batell}, by respecting the assumption that $U(1)_Y$
gauge boson is accessible to the extra dimension and the $SU(2)_L$
gauge bosons and the other particles, fermions, Higgs bosons, are
living on the visible brane, in the RS1 background. The starting
point is the 5D action
\begin{eqnarray}
S=\int d^5x \,\sqrt{-g}\,\Bigg\{-\frac{1}{4}\,F^{MN}\,F_{MN}+
\Bigg(-\frac{1}{4}\,G^{a\, \mu\nu}\,G^a_{\mu\nu}+ (D^\mu
\phi)^\dagger D_\mu \phi-V(\phi) \Bigg)\,\delta(y-\pi\, R) \Bigg\}
\,\, ,\label{Action2}
\end{eqnarray}
with field strength tensors $F^{MN}$ and $G^{a\, \mu\nu}$ for
$U(1)_Y$ and $SU(2)_L$ gauge bosons, where $M,N=0,...,4; \, \mu,
\nu=0,...,3$. The covariant derivative $D_\mu$ reads
\begin{eqnarray}
D_\mu=\partial_\mu-\frac{i}{2}\,g\,A_\mu^a
(x)\,\sigma^a-i\,g_5\,Y\,B_\mu (x,y)
 \,\, ,\label{Dmu}
\end{eqnarray}
where $B_\mu (x,y)$,  $A_\mu^a (x)$ are $U(1)_Y$, $SU(2)_L$ gauge
bosons, respectively; $g$, $g_5$ are the corresponding couplings
to the Higgs boson and $\sigma^a$ (Y=1/2) is the Pauli spin matrix
(the hypercharge). Here the gauge field $B_\mu (x,y)$ is expanded
to its Kaluza Klein (KK) modes as
\begin{eqnarray}
B_\mu (x,y)=\sum_n B_\mu^{(n)}(x)\,\chi^{(n)}(y)
 \,\, ,\label{KKsum}
\end{eqnarray}
and $\chi^{(n)}(y)$ satisfy the differential equation (see also
\cite{Pamoral,Hewett})
\begin{eqnarray}
(\frac{\partial^2}{\partial y^2}-2\,k\,\frac{\partial}{\partial
y}-m^2+e^{2\,k\,y}\, m_n^2)\, \chi^{(n)}(y)=0
 \,\, ,\label{diffeq}
\end{eqnarray}
with the normalization condition
\begin{eqnarray}
\int_0^{\pi\,R}\, dy\,\chi^{(n)}(y)\,\chi^{(m)}(y)=\delta_{nm}
\,\, . \label{ortagonal}
\end{eqnarray}
Here $k$ is the curvature scale and $m$ is the bulk mass of the
gauge field $B_\mu (x,y)$, which is at the order of the magnitude
of $k$. To obtain zero mode, which appears in the construction of
the photon and Z boson fields, the boundary mass term
\begin{eqnarray}
S'=-\int d^5x \,\sqrt{-g}\,\alpha\, k\,B^\mu (x,y)\, B_\mu (x,y)\,
\Big(\delta(y)-\delta(y-\pi\, R)\Big)\,\, ,\label{Action3}
\end{eqnarray}
is considered\footnote{The idea of brane localized mass terms has
been considered for scalar fields in
\cite{Goldberger},\cite{Pamoral2}.} and this term induces the
boundary condition
\begin{eqnarray}
\Big( \frac{\partial B_\mu (x,y)}{\partial y}-\alpha\,k\,B_\mu
(x,y)\Big)|_{y=0,\pi\,R}=0
 \,\, ,\label{Boundary}
\end{eqnarray}
which results in non-vanishing zero mode with the fine tuning of
the parameters $\alpha$ and $a=m^2/k^2$,
\begin{eqnarray}
\alpha_\pm=1\pm\sqrt{1+a}
 \,\, .\label{alfaa}
\end{eqnarray}
Finally, the normalized zero mode is obtained as
\begin{eqnarray}
\chi^{(0)}=\sqrt{\frac{2\,\alpha k}{e^{2\,\alpha \,
k\,\pi\,R}-1}}\, e^{\alpha\,k\,y} \,\, , \label{chi0}
\end{eqnarray}
and the $n$ mode reads
\begin{eqnarray}
\chi^{(n)}(y)=N_n\,
e^{k\,y}\,\Bigg(J_\nu(\frac{m_n}{k}\,e^{k\,y})+
b_{n\nu}\,Y_\nu(\frac{m_n}{k}\,e^{k\,y}) \Bigg) \,\, ,\label{chin}
\end{eqnarray}
with the normalization constant $N_n$, and the parameter
$\nu=\sqrt{1+a}$. Using the boundary condition at $y=0$ and
$y=\pi\,R$ (see eq.( \ref{Boundary})), the mass spectrum of KK
modes ($n=1,2,...$) reads
\begin{eqnarray}
m_n\simeq
(n\pm\frac{1}{2}\,\alpha_\pm-\frac{1}{4})\,\pi\,k\,e^{-k\,\pi\,R}
\,\, ,\label{mn}
\end{eqnarray}
for $k\,e^{-k\,\pi\,R}\ll m_n\ll k$.

Now, we present the mass Lagrangian for the photon and Z fields.
After allowing the Higgs boson vacuum expectation value (see eq.
(\ref{choice2})) and expanding the $U(1)_Y$ gauge field $B_\mu
(x,y)$ to KK modes (see eq. (\ref{KKsum}), the mass Lagrangian
becomes
\begin{eqnarray}
L_m=\sum_{n=1}^\infty \frac{1}{2}\,m_n^2\,
(B_\mu^{(n)}(x))^2+\frac{1}{2}\,(\frac{v}{2})^2\,\Bigg(
-g\,A_\mu^3(x)+g'\,B^0_\mu (x)+g'\,\sum_{n=1}^\infty
\frac{\chi^{(n)}(\pi\,R)}{\chi^{(0)}(\pi\,R)}\,B^n_\mu (x)
\Bigg)^2 \,\, ,\label{Lm}
\end{eqnarray}
where $g'=g_5\,\chi^{(0)}(\pi\,R)$. By considering the new basis
and using the mixing angle $\theta_W$, we rewrite the above
Lagrangian in terms of photon and Z fields as
\begin{eqnarray}
L_m=\sum_{n=1}^\infty  \frac{1}{2}\,m_n^2
(B_\mu^{(n)}(x))^2+\frac{1}{2}\,\Bigg(m_Z \,Z_\mu
(x)-\frac{g'\,v}{2}\,\sum_{n=1}^\infty
\frac{\chi^{(n)}(\pi\,R)}{\chi^{(0)}(\pi\,R)}\,B^n_\mu (x)
\Bigg)^2 \,\, .\label{Lm2}
\end{eqnarray}
Here the photon field is massless as it should be and there exists
mixing among Z boson and $B$ field KK modes. With the assumption
that $m_Z\ll m_n$, the diagonalization of the mass matrix results
in the shift in the mass of Z boson as
\begin{eqnarray}
m_Z^{phys}=m_Z\sqrt{1-(\frac{g'\,v}{2})^2\, \sum_{n=1}^\infty
\Big(\frac{1}{m_n}
\frac{\chi^{(n)}(\pi\,R)}{\chi^{(0)}(\pi\,R)}\Big)^2} \,\,
,\label{mZZ}
\end{eqnarray}
and the other physical masses coming from the mixing can be
approximated to $m_n$ (see eq. (\ref{mn})). On the other hand, the
coupling of zero mode (KK mode) physical Z boson to the fermions
reads $g'=g_5\, \chi^{(0)}$ ($g'^{(n)}\sim g_5\,\chi^{(n)})$. Here
$g'^{(n)}\sim g' \sqrt{\frac{1-e^{-2\alpha\,\pi\,k\,R}}{\alpha}}$
and, for positive $\alpha$, $g'^{(n)}\sim g'
\sqrt{\frac{1}{\alpha}}$. This is the case that the photon and Z
bosons are effectively localized around the visible brane.
\newpage
\section{Explicit Expresions}
The explicit expressions for the functions $f_1$, $f_2$, $f_3$ and
$f_4$ appearing in eq. (\ref{GammaRen2}) read
\begin{eqnarray}
f_1&=& \frac{g}{64\,\pi^2\,cos\,\theta_W} \int_0^1\, dx \,
\frac{1}{m^2_{l_2}-m^2_{l_1}} \Bigg \{ c_V \, (m_{l_2}+m_{l_1})
\nonumber \\
&\Bigg(& (-m_i \, \eta^+_i + m_{l_1} (-1+x)\, \eta_i^V)\, ln \,
\frac{L^{self}_ {1,\,h^0}}{\mu^2}+ (m_i \, \eta^+_i - m_{l_2}
(-1+x)\, \eta_i^V)\, ln \, \frac{L^{self}_{2,\, h^0}}{\mu^2}
\nonumber \\ &+& (m_i \, \eta^+_i + m_{l_1} (-1+x)\, \eta_i^V)\,
ln \, \frac{L^{self}_{1,\, A^0}}{\mu^2} - (m_i \, \eta^+_i +
m_{l_2} (-1+x) \,\eta_i^V)\, ln \, \frac{L^{self}_{2,\,
A^0}}{\mu^2} \Bigg) \nonumber \\ &+&
c_A \, (m_{l_2}-m_{l_1}) \nonumber \\
&\Bigg ( & (-m_i \, \eta^-_i + m_{l_1} (-1+x)\, \eta_i^A)\, ln \,
\frac{L^{self}_{1,\, h^0}}{\mu^2} + (m_i \, \eta^-_i + m_{l_2}
(-1+x)\, \eta_i^A)\, ln \, \frac{L^{self}_{2,\, h^0}}{\mu^2}
\nonumber \\ &+& (m_i \, \eta^-_i + m_{l_1} (-1+x)\, \eta_i^A)\,
ln \, \frac{L^{self}_{1,\, A^0}}{\mu^2} + (-m_i \, \eta^-_i +
m_{l_2} (-1+x)\, \eta_i^A)\, ln \, \frac{L^{self}_{2,\,
A^0}}{\mu^2} \Bigg) \Bigg \} \nonumber \\ &-&
\frac{g}{64\,\pi^2\,cos\,\theta_W} \int_0^1\,dx\, \int_0^{1-x} \,
dy \, \Bigg \{ m_i^2 \,(c_A\,
\eta_i^A-c_V\,\eta_i^V)\,(\frac{1}{L^{ver}_{A^0}}+
\frac{1}{L^{ver}_{h^0}}) \nonumber \\ &-& (1-x-y)\,m_i\, \Bigg(
c_A\,  (m_{l_2}-m_{l_1})\, \eta_i^- \,(\frac{1}{L^{ver}_{h^0}} -
\frac{1}{L^{ver}_{A^0}})+ c_V\, (m_{l_2}+m_{l_1})\, \eta_i^+ \,
(\frac{1}{L^{ver}_{h^0}} + \frac{1}{L^{ver}_{A^0}}) \Bigg)
\nonumber \\ &-& (c_A\, \eta_i^A+c_V\,\eta_i^V) \Bigg (
-2+(k^2\,x\,y+m_{l_1}\,m_{l_2}\, (-1+x+y)^2)\,
(\frac{1}{L^{ver}_{h^0}} +
\frac{1}{L^{ver}_{A^0}})-ln\,\frac{L^{ver}_{h^0}}{\mu^2}\,
\frac{L^{ver}_{A^0}}{\mu^2} \Bigg ) \nonumber \\ &-&
(m_{l_2}+m_{l_1})\, (1-x-y)\, \Bigg ( \frac{\eta_i^A\,(x\,m_{l_1}
+y\,m_{l_2})+m_i\,\eta_i^-}
{2\,L^{ver}_{A^0\,h^0}}+\frac{\eta_i^A\,(x\,m_{l_1} +y\,m_{l_2})-
m_i\,\eta_i^-}{2\,L^{ver}_{h^0\,A^0}} \Bigg ) \nonumber \\ &+&
\frac{1}{2}\eta_i^A\, ln\,\frac{L^{ver}_{A^0\,h^0}}{\mu^2}\,
\frac{L^{ver}_{h^0\,A^0}}{\mu^2}
\Bigg \}\,, \nonumber \\
f_2&=& \frac{g}{64\,\pi^2\,cos\,\theta_W} \int_0^1\, dx \,
\frac{1}{m^2_{l_2}-m^2_{l_1}} \Bigg \{ c_V \, (m_{l_2}-m_{l_1})
\nonumber \\
&\Bigg(& (m_i \, \eta^-_i + m_{l_1} (-1+x)\, \eta_i^A)\, ln \,
\frac{L^{self}_{1,\,A^0}}{\mu^2} + (-m_i \, \eta^-_i + m_{l_2}
(-1+x)\, \eta_i^A)\, ln \, \frac{L^{self}_ {2,\,A^0}}{\mu^2}
\nonumber \\ &+& (-m_i \, \eta^-_i + m_{l_1} (-1+x)\, \eta_i^A)\,
ln \, \frac{L^{self}_{1,\, h^0}}{\mu^2}+ (m_i \, \eta^-_i +
m_{l_2} (-1+x)\, \eta_i^A)\, ln \,
\frac{L^{self}_{2,\,h^0}}{\mu^2} \Bigg) \nonumber \\ &+&
c_A \, (m_{l_2}+m_{l_1}) \nonumber \\
&\Bigg(& (m_i \, \eta^+_i + m_{l_1} (-1+x)\, \eta_i^V)\, ln \,
\frac{L^{self}_{1,\, A^0}}{\mu^2}- (m_i \, \eta^+_i + m_{l_2}
(-1+x)\, \eta_i^V)\, ln \, \frac{L^{self}_{2,\,A^0}}{\mu^2}
\nonumber \\ &+& (-m_i \, \eta^+_i + m_{l_1} (-1+x)\, \eta_i^V)\,
ln \, \frac{L^{self}_{1,\, h^0}}{\mu^2} + (m_i \, \eta^+_i -
m_{l_2} (-1+x)\, \eta_i^V)\, \frac{ln \,
L^{self}_{2,\,h^0}}{\mu^2} \Bigg) \Bigg \} \nonumber \\ &+&
\frac{g}{64\,\pi^2\,cos\,\theta_W} \int_0^1\,dx\, \int_0^{1-x} \,
dy \, \Bigg \{ m_i^2 \,(c_V\,
\eta_i^A-c_A\,\eta_i^V)\,(\frac{1}{L^{ver}_{A^0}}+
\frac{1}{L^{ver}_{h^0}}) \nonumber \\ &-& m_i\, (1-x-y)\, \Bigg(
c_V\, (m_{l_2}-m_{l_1}) \,\eta_i^- + c_A\, (m_{l_2}+m_{l_1})\,
\eta_i^+ \Bigg) \,(\frac{1} {L^{ver}_{h^0}} -
\frac{1}{L^{ver}_{A^0}}) \nonumber \\ &+& (c_V\,
\eta_i^A+c_A\,\eta_i^V) \Bigg(-2+(k^2\,x\,y-m_{l_1}\,m_{l_2}\,
(-1+x+y)^2) (\frac{1}{L^{ver}_{h^0}}+\frac{1}{L^{ver}_{A^0}})-
ln\,\frac{L^{ver}_{h^0}}{\mu^2}\,\frac{L^{ver}_{A^0}}{\mu^2}
\Bigg) \nonumber \\ &-& (m_{l_2}-m_{l_1})\, (1-x-y)\, \Bigg(
\frac{\eta_i^V\,(x\,m_{l_1} -y\,m_{l_2})+m_i\,\eta_i^+}
{2\,L^{ver}_{A^0\,h^0}}+ \frac{\eta_i^V\,(x\,m_{l_1}
-y\,m_{l_2})-m_i\, \eta_i^+}{2\,L^{ver}_{h^0\,A^0}}
\Bigg)\nonumber \\
&-& \frac{1}{2} \eta_i^V\, ln\,\frac{L^{ver}_{A^0\,h^0}}{\mu^2}\,
\frac{L^{ver}_{h^0\,A^0}}{\mu^2}
\Bigg \} \nonumber \,,\\
f_3&=&-i \frac{g}{64\,\pi^2\,cos\,\theta_W} \int_0^1\,dx\,
\int_0^{1-x} \, dy \, \Bigg \{ \Bigg( (1-x-y)\,(c_V\,
\eta_i^V+c_A\,\eta_i^A)\, (x\,m_{l_1} +y\,m_{l_2}) \nonumber
\\ &+& \, m_i\,(c_A\, (x-y)\,\eta_i^-+c_V\,\eta_i^+\,(x+y))\Bigg )
\,\frac{1}{L^{ver}_{h^0}} \nonumber \\ &+& \Bigg( (1-x-y)\, (c_V\,
\eta_i^V+c_A\,\eta_i^A)\, (x\,m_{l_1} +y\,m_{l_2}) -m_i\,(c_A\,
(x-y)\,\eta_i^-+c_V\,\eta_i^+\,(x+y))\Bigg )
\,\frac{1}{L^{ver}_{A^0}} \nonumber \\ &-& (1-x-y) \Bigg
(\frac{\eta_i^A\,(x\,m_{l_1} +y\,m_{l_2})}{2}\, \Big (
\frac{1}{L^{ver}_{A^0\,h^0}}+\frac{1}{L^{ver}_{h^0\,A^0}} \Big )
+\frac{m_i\,\eta_i^-} {2} \, \Big ( \frac{1}{L^{ver}_{h^0\,A^0}}-
\frac{1}{L^{ver}_{A^0\,h^0}} \Big ) \Bigg ) \Bigg \} \,,\nonumber \\
f_4&=&-i\frac{g}{64\,\pi^2\, cos\,\theta_W} \int_0^1\,dx\,
\int_0^{1-x} \, dy \, \Bigg \{ \Bigg( (1-x-y)\,\Big ( -(c_V\,
\eta_i^A+c_A\,\eta_i^V)\, (x\,m_{l_1} -y\, m_{l_2}) \Big)
\nonumber \\ &-& m_i\, (c_A\,
(x-y)\,\eta_i^++c_V\,\eta_i^-\,(x+y))\Bigg )\,
\frac{1}{L^{ver}_{h^0}} \nonumber \\ &+& \Bigg ( (1-x-y)\,\Big (
-(c_V\, \eta_i^A+c_A\,\eta_i^V)\, (x\,m_{l_1} - y\, m_{l_2}) \Big
) + m_i\,(c_A\, (x-y)\,\eta_i^++c_V\,\eta_i^-\,(x+y)) \Bigg )
\,\frac{1}{L^{ver}_{A^0}} \nonumber \\&+& (1-x-y)\, \Bigg (
\frac{\eta_i^V}{2}\,(m_{l_1}\,x-m_{l_2}\, y)\, \, \Big (
\frac{1}{L^{ver}_{A^0\,h^0}}+\frac{1}{L^{ver}_{h^0\,A^0}} \Big )
+\frac{m_i\,\eta_i^+}{2}\, \Big (
\frac{1}{L^{ver}_{A^0\,h^0}}-\frac{1}{L^{ver}_{h^0\,A^0}} \Big )
\Bigg ) \Bigg \}\, , \label{fVAME}
\end{eqnarray}
where
\begin{eqnarray}
L^{self}_{1,\,h^0}&=&m_{h^0}^2\,(1-x)+(m_i^2-m^2_{l_1}\,(1-x))\,x
\nonumber \, , \\
L^{self}_{1,\,A^0}&=&L^{self}_{1,\,h^0}(m_{h^0}\rightarrow
m_{A^0})
\nonumber \, , \\
L^{self}_{2,\,h^0}&=&L^{self}_{1,\,h^0}(m_{l_1}\rightarrow
m_{l_2})
\nonumber \, , \\
L^{self}_{2,\,A^0}&=&L^{self}_{1,\,A^0}(m_{l_1}\rightarrow
m_{l_2})
\nonumber \, , \\
L^{ver}_{h^0}&=&m_{h^0}^2\,(1-x-y)+m_i^2\,(x+y)-k^2\,x\,y
\nonumber \, , \\
L^{ver}_{h^0\,A^0}&=&m_{A^0}^2\,x+m_i^2\,(1-x-y)+(m_{h^0}^2-k^2\,x)\,y
\nonumber \, , \\
L^{ver}_{A^0}&=&L^{ver}_{h^0}(m_{h^0}\rightarrow m_{A^0})
\nonumber \, , \\
L^{ver}_{A^0\,h^0}&=&L^{ver}_{h^0\,A^0}(m_{h^0}\rightarrow
m_{A^0}) \, , \label{Lh0A0}
\end{eqnarray}
and
\begin{eqnarray}
\eta_i^V&=&\xi^{E}_{N,l_1i}\xi^{E\,*}_{N,il_2}+
\xi^{E\,*}_{N,il_1} \xi^{E}_{N,l_2 i} \nonumber \, , \\
\eta_i^A&=&\xi^{E}_{N,l_1i}\xi^{E\,*}_{N,il_2}-
\xi^{E\,*}_{N,il_1} \xi^{E}_{N,l_2 i} \nonumber \, , \\
\eta_i^+&=&\xi^{E\,*}_{N,il_1}\xi^{E\,*}_{N,il_2}+
\xi^{E}_{N,l_1i} \xi^{E}_{N,l_2 i} \nonumber \, , \\
\eta_i^-&=&\xi^{E\,*}_{N,il_1}\xi^{E\,*}_{N,il_2}-
\xi^{E}_{N,l_1i} \xi^{E}_{N,l_2 i}\, . \label{etaVA}
\end{eqnarray}
The parameters $c_V$ and $c_A$ are $c_A=-\frac{1}{4}$ and
$c_V=\frac{1}{4}-sin^2\,\theta_W$. In eq. (\ref{etaVA}) the flavor
changing couplings $\xi^{E}_{N, ji}$ represent the effective
interaction between the internal lepton $i$, ($i=e,\mu,\tau$) and
outgoing (incoming) $j=l_1\,(j=l_2)$ one. Here we take only the
$\tau$ lepton in the internal line and  we neglect all the Yukawa
couplings except $\xi_{N,\tau\tau}^E$ and $\xi_{N,\tau\mu}^E$ in
the loop contributions (see Discussion section). The  Yukawa
couplings $\xi^{E}_{N, ji}$ are complex in general, however, in
the present work, we take them real.
\newpage
\newpage
\end{document}